\documentclass[12pt]{article}
\usepackage{epsfig}
\def\simlt{\stackrel{<}{{}_\sim}}
\def\simgt{\stackrel{>}{{}_\sim}}
\def\NPB#1#2#3{{\it Nucl.~Phys.} {\bf{B#1}} (#2) #3}
\def\PLB#1#2#3{{\it Phys.~Lett.} {\bf{B#1}} (#2) #3}
\def\PRD#1#2#3{{\it Phys.~Rev.} {\bf{D#1}} (#2) #3}
\def\PRL#1#2#3{{\it Phys.~Rev.~Lett.} {\bf{#1}} (#2) #3}
\def\ZPC#1#2#3{{\it Z.~Phys.} {\bf C#1} (#2) #3}

\def\be{\begin{equation}}
\def\ee{\end{equation}}
\def\bea{\begin{eqnarray}}
\def\eea{\end{eqnarray}}
\def\beq{\begin{equation}}
\def\eeq{\end{equation}}
\def\bq{\begin{quote}}
\def\eq{\end{quote}}

\def\ie{{\it i.e.}}

\def\frac#1#2{{{#1}\over {#2}}}

\def\bq{\bar{q}}
\catcode`@=11 
\def\slash#1{\mathord{\mathpalette\c@ncel#1}}
\def\c@ncel#1#2{\ooalign{$\hfil#1\mkern1mu/\hfil$\crcr$#1#2$}}
\def\lsim{\mathrel{\mathpalette\@versim<}}
\def\gsim{\mathrel{\mathpalette\@versim>}}
 \def\@versim#1#2{\lower0.2ex\vbox{\baselineskip\z@skip\lineskip\z@skip
       \lineskiplimit\z@\ialign{$\m@th#1\hfil##$\crcr#2\crcr\sim\crcr}}}
\catcode`@=12 

\newpage
\begin{document} 
\pagestyle{empty} 
\begin{flushright}
{\tt hep-ph/0106029}\\
CERN-TH/2001-145\\
GeF/TH/8-01\\
IFUM-689-FT
\end{flushright}
\vspace*{5mm}
\begin{center}
{\large\bf Indication for Light Sneutrinos and Gauginos from Precision
Electroweak Data}\\
\vspace*{0.8cm} {\bf G.~Altarelli,$^a$
F.~Caravaglios,$^b$ G.F.~Giudice,$^a$\\
P.~Gambino,$^a$ G.~Ridolfi$^{a,c}$}
\\
\vspace{0.6cm} 
$^a$ Theoretical Physics Division, CERN, CH--1211 Geneva 23, Switzerland. \\
$^b$ Dipartimento di Fisica, Universit\`a di Milano, Via Celoria 16,
I--20133 Milan, Italy.\\
$^c$ On leave from INFN, Sezione di Genova, Via Dodecaneso 33, I--16146 Genoa,
Italy.

\vspace*{0.9cm} {\bf Abstract}
\end{center}
\noindent
The present Standard Model fit of precision data has a low confidence
level, and is characterized by a few inconsistencies. We look for
supersymmetric effects that could improve the agreement among the
electroweak precision measurements and with the direct lower bound on
the Higgs mass. We find that this is the case particularly if the
$3.6~\sigma$ discrepancy between $\sin^2\theta_{\rm eff}$ from
leptonic and hadronic asymmetries is finally settled more on the side
of the leptonic ones. After the inclusion of all experimental
constraints, our analysis selects light sneutrinos, with masses in the
range $55-80$~GeV, and charged sleptons with masses just above their
experimental limit, possibly with additional effects from light
gauginos. The phenomenological implications of this scenario are
discussed.

\vfill
\begin{flushleft} CERN-TH/2001-145\\ June 2001 \end{flushleft} 
\eject 

\setcounter{page}{1} \pagestyle{plain}
\section{Introduction}
The results of the electroweak precision tests as well as of the
searches for the Higgs boson and for new particles performed at LEP
and SLC have now been presented in a close to final form.  Taken
together with the measurements of $m_t$, $m_W$ and the searches for
new physics at the Tevatron, and with some other data from low energy
experiments, they form a very stringent set of precise constraints to
compare with the Standard Model (SM) or with any of its conceivable
extensions. When confronted with these results, on the whole the SM
performs rather well, so that it is fair to say that no clear
indication for new physics emerges from the data.  However, if we look
at the results in detail, there are a number of features that are
either not satisfactory or could indicate the presence of small new
physics effects. We will describe in quantitative terms the
experimental results and their consistency among themselves and with
the SM in the next section. Here we anticipate a qualitative
discussion.

One problem is that the two most precise measurements of
$\sin^2\theta_{\rm eff}$ from $A_{LR}$ and $A^b_{FB}$ differ by
$3.5~\sigma$~\cite{EWWG}. More in general, there appears to be a
discrepancy between $\sin^2\theta_{\rm eff}$ measured from leptonic
asymmetries and from hadronic asymmetries. The result from $A_{LR}$ is
actually in good agreement with the leptonic asymmetries measured at
LEP, while all hadronic asymmetries are better compatible with the
result of $A^b_{FB}$. It is well known that this discrepancy is not
likely to be explained by some new physics effect in the $b \bar b Z$
vertex. In fact $A^b_{FB}$ is the product of lepton- and $b$-asymmetry
factors: $A^b_{FB}\propto A_\ell A_b$, where
$A_f=2g^f_Ag^f_V/({g^f_A}^2+{g^f_V}^2)$. The sensitivity of
$A^b_{FB}$ to $A_b$ is limited, because the $A_\ell$ factor is small,
so that, in order to reproduce the measured
discrepancy, the new effect should induce a large change of the $b$
couplings with respect to the SM.
But then this effect should be clearly visible in the
direct measurement of $A_b$ performed at SLD using the LR polarized
$b$ asymmetry, even within the moderate precision of this result, and
it should also appear in the accurate measurement of $R_b\propto
{g^{b}_A}^2+{g^{b}_V}^2$. Neither $A_b$ nor $R_b$ show deviations of
the expected size. One concludes that most probably the observed
discrepancy is due to a large statistical fluctuation and/or to an
experimental problem. Indeed, the measurement of $A^b_{FB}$ not only
requires $b$ identification, but also distinguishing $b$ from
$\bar{b}$, and therefore the systematics involved are different than
in the measurement of $R_b$. At any rate, the disagreement between
$A^b_{FB}$ and $A_{LR}$ implies that the ambiguity in the measured
value of $\sin^2\theta_{\rm eff}$ is larger than the nominal error
obtained from averaging all the existing determinations.

Another point of focus is the relation between the fitted Higgs mass
and the lower limit on this mass from direct searches, $m_H >
113$~GeV, as it was recently stressed in ref.~\cite{cha}. The central
value of the fitted mass is systematically below the limit. In
particular, given the experimental value of the top mass, the measured
results for $m_W$ (with perfect agreement between LEP and the
Tevatron) and $\sin^2\theta_{\rm eff}$ measured from leptonic
asymmetries, taken together with the results on the $Z_0$ partial
widths, push the central value of $m_H$ very much down. In fact, if
one arbitrarily excludes $\sin^2\theta_{\rm eff}$ measured from the
hadronic asymmetries, the fitted value of $m_H$ becomes only
marginally consistent with the direct limit, to a level that depends
on the adopted value and the error for $\alpha_{QED}(m_Z)$.
Consistency is reinstated if the results from hadronic asymmetries are
also included, because they drive the fitted $m_H$ value towards
somewhat larger values.

In conclusion, if one takes all available measurements into account
the $\chi^2$ of the SM fit is not good, with a probability of about
$4\%$, partly because the measurements of $\sin^2\theta_{\rm eff}$ are
not in good agreement among them. If, on the other hand, one only
takes the results on $\sin^2\theta_{\rm eff}$ from the leptonic
asymmetries, then the $\chi^2$ of the SM fit considerably improves,
but the consistency with the direct limit on $m_H$ becomes marginal.

In this article we enlarge the discussion of the data from the SM to
the Minimal Supersymmetric Standard Model (MSSM). We look for regions
of the MSSM parameter space where the corrections are sufficiently large
and act in the direction of improving
the quality of the fit and the
consistency with the direct limit on $m_H$ with respect to the
SM, especially in the most unfavourable case for the SM that the
results on $\sin^2\theta_{\rm eff}$ from the hadronic asymmetries are
discarded.  We will show that, if sleptons (and, to a lesser extent,
charginos and neutralinos) have masses close to their present
experimental limits, it is possible to considerably improve the overall
picture. In particular the possible MSSM effects become sizeable if we
allow the sneutrino masses to be as small as allowed by the direct
limits on $m^2_{\tilde{\nu}}$ and by those on charged slepton masses,
which are related by $m^2_{\tilde{\ell}^\pm_L}=m^2_{\tilde\nu}+m^2_W|\cos
2\beta|$.  At moderately large values of $\tan\beta$ (\ie\ for $|\cos
2\beta|\sim 1$), light sneutrinos with masses as low as $55$~GeV
are not excluded by present limits, while charged sleptons must be
heavier than $96$~GeV. These low values of the sneutrino mass
can still be compatible with the neutralino being the lightest
supersymmetric particle. This region of parameter space was not
emphasized in some past analyses~\cite{MSSMstudies,Pierce,Erler}.
We recall that $\tan\beta\simgt 2-3$ is
required by LEP, and large $\tan\beta$ and light sleptons are
indicated by the possible deviation observed by the recent Brookhaven
result~\cite{gminustwo} on the muon $g-2$, if this discrepancy is to be
explained by a MSSM effect. We find it interesting that, by taking
seriously the small hints that appear in the present data, one can
pinpoint a region of the MSSM which match the data better than the SM,
and is likely to be within reach of the present run of the Tevatron and, of
course, of the LHC.

For this analysis in the MSSM we use the technique of the epsilon
parameters $\epsilon_1$, $\epsilon_2$, $\epsilon_3$ and $\epsilon_b$,
introduced in ref.~\cite{epsilon}.  The variations of $\epsilon_1$,
$\epsilon_2$ and $\epsilon_3$ due to new physics contributions are
proportional to the shifts in the $T$, $U$, and $S$
parameters~\cite{Peskin}, respectively, if one keeps only oblique
contributions (\ie\ terms arising from vacuum polarization diagrams),
expanded up to the first power in the external momentum squared.  But
in the MSSM not all important contributions are of this
kind~\cite{Erler}.  We recall that the starting point of the epsilon
analysis is the unambiguous definition of the $\epsilon_i$ in terms of
four basic observables that were chosen to be $\sin^2\theta_{\rm eff}$
from $A^\mu_{FB}$, $\Gamma_\mu$, $m_W$ and $R_b$. Given the
experimental values of these quantities, the corresponding
experimental values of the $\epsilon_i$ follow, independent of $m_t$
and $m_H$, with an error that, in addition to the propagation of the
experimental errors, also includes the effect of the present
ambiguities in $\alpha_s(m_Z)$ and $\alpha_{QED}(m_Z)$.

If one assumes lepton universality, which is well supported by the
data within the present accuracy, then the combined results on
$\sin^2\theta_{\rm eff}$ from all leptonic asymmetries can be adopted
together with the combined leptonic partial width $\Gamma_\ell$. At
this level the epsilon analysis is model-independent within the stated
lepton universality assumption. As a further step we can observe that
by including the information on the hadronic widths arising from
$\Gamma_Z$, $\sigma_h$, $R_\ell$, the central values of the
$\epsilon_i$ are not much changed (with respect to the error size) and
the errors are slightly decreased.  Thus one may decide of including
or not including these data in the determination of the $\epsilon_i$,
without affecting the results.

Different is the case of including the results from the hadronic
asymmetries in the combined value of $\sin^2\theta_{\rm eff}$. In this
case, obviously, the determination of $\epsilon_i$ is sizeably
affected and one remains with the alternative between an experimental
problem or a bizarre effect of some new physics in the $b$ coupling
(not present in the MSSM). But if we remain within the first stage of
purely leptonic measurements plus $m_W$ and $R_b$, the $\epsilon_i$
analysis is quite general and, in particular, is independent of an
assumption of oblique correction dominance.

The comparison with the SM can be repeated in the context of the
$\epsilon_i$. The predicted theoretical values of the $\epsilon_i$ in
the SM depend on $m_H$ and $m_t$, while they are practically
independent of $\alpha_s(m_Z)$ and $\alpha_{QED}(m_Z)$. If we only
take the leptonic measurements of $\sin^2\theta_{\rm eff}$, for
$m_H=113$~GeV and $m_t=174.3$~GeV one finds that the experimental value
of $\epsilon_1$ agrees within the error with the prediction, while
both $\epsilon_2$ and $\epsilon_3$ are below the theoretical
expectation by about $1~\sigma$. We recall that $m_W$ is
related to $\epsilon_2$ and the fact that the experimental value is
below the prediction for this quantity corresponds to the statement
that $m_W$ would prefer a value of $m_H$ much smaller than
$m_H=113$~GeV. Similarly the smallness of the fitted value of
$\epsilon_3$ with respect to the prediction has to do with the marked
preference for a light $m_H$ of $\sin^2\theta_{\rm eff}$ from all
leptonic asymmetries. The agreement between fitted value and
prediction for $\epsilon_1$, which, contrary to $\epsilon_2$ and
$\epsilon_3$, contains a quadratic dependence on $m_t$, reflects the
fact that the fitted value of $m_t$ is in agreement with the
measured value. The other variable that depends quadratically on $m_t$
is $\epsilon_b$. The agreement of the fitted and predicted values of
$\epsilon_b$ reflects the corresponding present normality of the
results for $R_b$.

\section{The data and their comparison with the Standard Model}
We start by summarising the different existing determinations of
$\sin^2\theta_{\rm eff}$ and their mutual consistency. The two most
precise measurements from $A_{LR}$ by SLD and $A^b_{FB}$ by LEP lead
to
\bea
&&\sin^2\theta_{\rm eff}~=~0.23098 \pm 0.00026\qquad\rm{(A_{LR})}\label{alr}
\\
&&\sin^2\theta_{\rm eff}~=~0.23240 \pm 0.00031\qquad\rm{(A^b_{FB})}.
\label{abfb}
\eea
As already mentioned these two measurements differ by $3.5\sigma$. If
we take $\sin^2\theta_{\rm eff}$ from the combined LEP/SLD leptonic or
hadronic asymmetries we have
\bea
&&\sin^2\theta_{\rm eff}~=~0.23114\pm0.00020\qquad
\rm{(all~leptonic~asymmetries)}\label{alep}\\
&&\sin^2\theta_{\rm eff}~=~0.23240\pm0.00029\qquad
\rm{(all~hadronic~asymmetries)}.
\label{ahad}
\eea
The resulting discrepancy is at $3.6~\sigma$, thus at about the same
level. By combining all the above measurements one obtains
\beq
\sin^2\theta_{\rm eff}~=~0.23156\pm0.00017\qquad
\rm{(all~asymmetries)}.\label{aall}
\eeq
We see that the dispersion between the results from leptonic and
hadronic asymmetries is much larger than the nominal error in the
combination.

The experimental values~\cite{EWWG} of the most important electroweak
observables which are used in our analysis are collected in table~\ref{tab1}.

\begin{table}[t]
\begin{center} 
\begin{tabular}{|l|c|}
\hline Quantity&Data (March 2001)     \\
\hline
  $m_Z$ (GeV)             & 91.1875(21)   \\
  $\Gamma_Z$ (GeV)        &  2.4952(23)   \\
  $\sigma_h$ (nb)         & 41.540(37)    \\
  $R_\ell$                & 20.767(25)    \\
  $R_b$                   &  0.21664(68)  \\
  $\Gamma_\ell$ (MeV)     & 83.984(86)    \\
  $A^\ell_{FB}$           &  0.01714(95)  \\
  $A_\tau$                &  0.1439(41)   \\
  $A_e$                   &  0.1498(48)   \\
  $A^b_{FB}$              &  0.0982(17)   \\
  $A^c_{FB}$              &  0.0689(35)   \\
  $A_b$ (SLD direct)      &  0.921(20)    \\ 
  $\sin^2\theta_{\rm eff}$ (all lept. asym.) & 0.23114(20) \\
  $\sin^2\theta_{\rm eff}$ (all hadr. asym.) & 0.23240(29) \\
  $m_W$ (GeV) (LEP2+$p\bar p$)               & 80.448(34)  \\
  $m_t$ (GeV)             & 174.3(5.1)  \\
  $\alpha_s(m_Z)$         &   0.119(3)  \\
\hline
\end{tabular}
\caption{\label{tab1}\it Observables included in our global fit.}
\end{center}
\end{table}
A quantity which plays a very important role in the interpretation of
the electroweak precision tests is the value of $\alpha_{QED}(m_Z)$,
the QED coupling at the scale $m_Z$ or, equivalently,
$\Delta\alpha_h$, the hadronic contribution to the shift
$\Delta\alpha$, with $\alpha_{QED}(m_Z)=\alpha/(1-\Delta\alpha)$. We
adopt here as our main reference values those recently derived in
ref.~\cite{BP01}:
\beq
\Delta\alpha_h=0.02761\pm 0.00036,
\qquad\alpha^{-1}_{QED}(m_Z)=128.936\pm0.049
\qquad\rm{(BP01)}.\label{bp}
\eeq 
A larger set of recent determinations of $\Delta\alpha_h$ will also be
used for comparison (see table~\ref{tab2}).

We consider now different SM fits to the observables $m_t$, $m_W$, 
$\Gamma_\ell$, $R_b$, $\alpha_s(m_Z)$, $\alpha_{QED}$, with
different assumptions on the input value of $\sin^2\theta_{\rm eff}$.
These fits are based on up-to-date theoretical calculations~\cite{degrassi}.
We start by considering $\sin^2\theta_{\rm eff}$ from all leptonic asymmetries,
eq.~(\ref{alep}),
and $\sin^2\theta_{\rm eff}$ from all hadronic asymmetries, eq.~(\ref{ahad}),
as two distinct inputs in the same fit. In this case,
we find $\chi^2/{\rm d.o.f.}=18.4/4$, corresponding to 
C.L.=$0.001$. When a more complete analysis is performed, including
all 20 observables in the global fit,
the situation is still not satisfying, although less dramatic:
ref.~\cite{EWWG} reports $\chi^2/{\rm d.o.f.}=26/15$,
with C.L.=$0.04$. If we now exclude $\sin^2\theta_{\rm eff}$
from all hadronic asymmetries, the quality of the fit
of our seven observables significantly improves, giving
$\chi^2/{\rm d.o.f.}=2.5/3$, C.L.=$0.48$,
while the fit to all observables except $A^b_{FB}$ gives~\cite{cha}
$\chi^2/{\rm d.o.f.}=15.8/14$, C.L.=$0.33$.
Finally, if we instead exclude $\sin^2\theta_{\rm eff}$ from all leptonic
asymmetries, we find $\chi^2/{\rm d.o.f.}=15.3/3$, C.L.=$0.0016$. Thus it
appears that the leptonic value of $\sin^2\theta_{\rm eff}$ 
leads to the best overall consistency in terms of C.L..

We now consider the corresponding fitted values of the Higgs mass, and
the 95\% C.L. upper limits. In the first case studied above, namely
when $\sin^2\theta_{\rm eff}$ from both hadronic and leptonic
asymmetries are included, with $\Delta\alpha_h^{\rm BP01}$ given in
eq.~(\ref{bp}), we obtain a central value for the Higgs mass of
$m_H=100$~GeV, with a $95\%$ C.L. limit $m_H\simlt 212$~GeV. These
values are indeed in complete agreement with the SM fit results
presented by the LEP Electroweak Group~\cite{EWWG},
based on the complete set of observables:
$m_H=98$~GeV and $m_H\simlt 212$~GeV. Neglecting the fact that the
dispersion of the various measurements corresponds to a very poor
$\chi^2$, there is no significant contradiction with the direct limit
on $m_H$. However, it is well known and was recently emphasized in
ref.~\cite{cha} that, if instead we use $\sin^2\theta_{\rm eff}$
measured from leptonic asymmetries only, see eq.~(\ref{alep}), which leads
to the best value of $\chi^2/{\rm d.o.f.}$, then the fitted value of
$m_H$ markedly drops and the consistency with the direct limit becomes
marginal.  In fact, in this case, all other inputs being the same, we
find $m_H=42$~GeV and $m_H\simlt 109$~GeV. In table~\ref{tab2} we
report the corresponding results for some other determinations of
$\Delta\alpha_h$. We see that, while there is some sensitivity to this
choice, the conclusion that the compatibility of the fitted value of
$m_H$ with the direct limit becomes marginal is quite stable.
Similarly, we believe that uncalculated higher order effects cannot
have a serious impact, as they can be estimated~\cite{barcelona} to
shift the 95\% C.L. up by at most 10-15 GeV.

It must however be recalled that the level of compatibility is
sensitive to the top mass, and is increased if $m_t$ is moved up within
its error bar: for a shift up by $1\sigma$ we find, using
$\Delta\alpha_h^{\rm BP01}$, $m_H=58$~GeV and $m_H\simlt 156$~GeV.

\begin{table}
\begin{center} 
\begin{tabular}{|l|c|c|c|}
\hline Ref. & $\Delta\alpha_h$ & $m_H$ (GeV) & $95\%$ C.L. limit (GeV) \\
\hline
BP01 \cite{BP01}      & 0.02761(36)   & 42  & 109 \\
J01 \cite{J01}        & 0.027896(395) & 34  &  91 \\
Jeucl01 \cite{J01}    & 0.027730(209) & 40  &  98 \\
MOR00 \cite{MOR00}    & 0.02738(20)   & 52  & 124 \\
DH98 \cite{DH98}      & 0.02763(16)   & 42  & 104 \\
KS98 \cite{KS98}      & 0.02775(17)   & 38  &  96 \\
EJ95 \cite{EJ95}      & 0.02804(65)   & 28  &  90 \\
\hline
\end{tabular}
\caption{\label{tab2}\it Different determinations of $\Delta\alpha_h$ and their
influence on the fitted Higgs mass.}
\end{center}
\end{table}

We now consider the epsilon analysis. As already mentioned,
the predicted values of the epsilon variables in the SM depend on $m_t$ and
$m_H$, while they are practically insensitive to small variations of
$\alpha_s(m_Z)$ and $\alpha_{QED}(m_Z)$. We report here the values of
$\epsilon_i$ for $m_H=113$~GeV and $m_t=174.3-5.1, 174.3,
174.3+5.1$~GeV, respectively:
\be
\begin{array}{cccc}
\epsilon_1= &~~5.1\times 10^{-3}, &~~5.6\times 10^{-3}, &~~6.0\times 10^{-3} \\
\epsilon_2= & -7.2\times 10^{-3}, & -7.4\times 10^{-3}, & -7.6\times 10^{-3} \\
\epsilon_3= &~~5.4\times 10^{-3}, &~~5.4\times 10^{-3}, &~~5.3\times 10^{-3} \\
\epsilon_b= & -6.2\times 10^{-3}, & -6.6\times 10^{-3}, & -7.1\times 10^{-3}.\\
\end{array}
\ee
We first consider the
observables $\sin^2\theta_{\rm eff}$ measured from leptonic
asymmetries, see eq.~(\ref{alep}), $\Gamma_\ell$, $m_W$, and $R_b$. From
these observables we obtain the following values of the $\epsilon_i$:
\be
\begin{array}{c}
\epsilon_1=~~(5.1\pm  1.0)\times 10^{-3}\\
\epsilon_2=(-9.0\pm 1.2)\times 10^{-3}\\
\epsilon_3=~~(4.2\pm  1.0)\times 10^{-3}\\
\epsilon_b=(-4.2\pm 1.8)\times 10^{-3}.\label{epsa}
\end{array}
\ee
(in this last fit, the value of $\alpha_s$ was kept fixed).
The errors also include the effect of the quoted errors on
$\alpha_s(m_Z)$ and $\alpha_{QED}(m_Z)$.  At this stage we have only
assumed lepton universality and the derivation of the $\epsilon_i$ is
otherwise completely model independent. For example, no assumption of
oblique corrections dominance is to be made. It is interesting to
observe that if we add to the previous set of observables the
information on the hadronic widths arising from $\Gamma_Z$,
$\sigma_h$, $R_\ell$ we obtain for the $\epsilon_i$
\be
\begin{array}{c}
\epsilon_1=~~(5.0\pm1.0)\times 10^{-3}\\
\epsilon_2=(-9.1\pm1.2)\times 10^{-3}\\
\epsilon_3=~~(4.2\pm1.0)\times 10^{-3}\\
\epsilon_b=(-5.7\pm1.6)\times 10^{-3}.\label{epsb}
\end{array}
\ee
The central values are only changed by a small amount (in comparison
with the error size) with respect to the previous fit. We interpret
this result by concluding that the hadronic $Z_0$ widths are perfectly
compatible with the leptonic widths. Thus, if there are new physics
corrections in the widths, these must be mostly of universal type like
from vacuum polarization diagrams.  A posteriori we can add this
information in the epsilon analysis which allows to slightly reduce
the errors on the individual $\epsilon_i$.

We can now consider the sensitivity of the $\epsilon_i$ to the
different determinations of $\sin^2\theta_{\rm eff}$. We take the same
set of observables as in the previous fit in eqs.~(\ref{epsb}), but
replace $\sin^2\theta_{\rm eff}$ from leptonic asymmetries with that
obtained from all combined measurements, as given in eq.~(\ref{aall}).
The corresponding values of the $\epsilon_i$ are given by
\be
\begin{array}{c}
\epsilon_1=~~(5.4\pm1.0)\times 10^{-3}\\
\epsilon_2=(-9.7\pm1.2)\times 10^{-3}\\
\epsilon_3=~~(5.4\pm0.9)\times 10^{-3}\\
\epsilon_b=(-5.5\pm1.6)\times 10^{-3}.\label{epsc}
\end{array}
\ee
We see that the most sensitive variable to $\sin^2\theta_{\rm eff}$ is
$\epsilon_3$ that changes by more than $1\sigma$ in the direction of a
better agreement with the SM prediction for $m_H=113$~GeV, but the
value of $\epsilon_2$ is even further away from the SM prediction.
This is in agreement with the results obtained in the direct analysis
of the data in the SM.

\begin{figure}[htb]
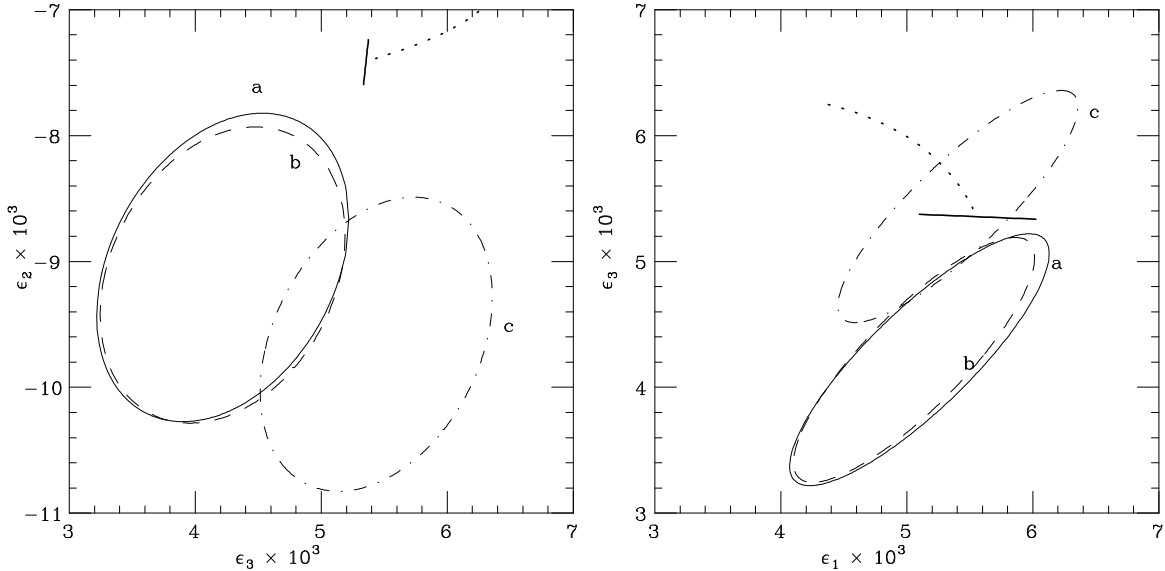

\begin{center}
\epsfig{file=fig1-l.ps,width=0.45\textwidth}
\epsfig{file=fig1-r.ps,width=0.45\textwidth}
\caption{\label{fig1}\it
One-sigma ellipses in the $\epsilon_3-\epsilon_2$ (left)
and in the $\epsilon_1-\epsilon_3$ (right) planes obtained from:
{\rm a}.~$m_W$, $\Gamma_\ell$, $\sin^2\theta_{\rm eff}$ from all leptonic
asymmetries, and $R_b$; {\rm b}.~the same observables, plus the hadronic 
partial widths derived from $\Gamma_Z$, $\sigma_h$ and $R_\ell$; 
{\rm c}.~as in {\rm b}.,
but with $\sin^2\theta_{\rm eff}$ also including the hadronic
asymmetry results. The solid straight lines represent
the SM predictions for $m_H=113$~{\rm GeV}
and $m_t$ in the range $174.3\pm 5.1$~{\rm GeV}. The dotted curves represent
the SM predictions for $m_t=174.3$~{\rm GeV} and $m_H$ in the range $113$ to 
$500$~{\rm GeV}.}
\end{center}
\end{figure}

The results of the above fits of the $\epsilon_i$, including the error
correlations among different variables, are shown in
fig.~\ref{fig1}. In these figures we display the $1\sigma$ ellipses
in the $\epsilon_i$-$\epsilon_j$ plane that correspond to the fits in
eqs.~(\ref{epsa},\ref{epsb},\ref{epsc}).  Note that these ellipses
project $\pm 1\sigma$ errors on either axis.  As such the probability
for both $\epsilon_i$ and $\epsilon_j$ to fall inside the ellipse is
only about $39\%$. The ellipses that correspond to other significance
levels can be obtained by scaling the ellipse axes by suitable well
known factors. We note the following salient features. The fitted
values of $\epsilon_1$ are in all cases perfectly compatible with the
predicted value in the SM. This corresponds to the fact that the
fitted and the measured values of $m_t$ coincide. The fitted values of
$\epsilon_2$ are always below the prediction, reflecting the fact that
the measured value of $m_W$ would prefer smaller $m_H$ and/or larger
$m_t$. The $\epsilon_2$ deviation is larger when also the measurement
of $\sin^2\theta_{\rm eff}$ from the hadronic asymmetries is included.
The fitted values of $\epsilon_3$ are below the prediction if the
value of $\sin^2\theta_{\rm eff}$ from leptonic asymmetries is used,
while the agreement is restored if the measurement of
$\sin^2\theta_{\rm eff}$ from the hadronic asymmetries is included.

In conclusion, the epsilon analysis reproduces the results obtained
from the direct comparison of the data with the SM. The most important
features are that both $m_W$ and $\sin^2\theta_{\rm eff}$ from
leptonic asymmetries appear to favour small $m_H$ and/or large $m_t$.
In the following we will discuss the effect of supersymmetry and the
choice of MSSM parameters that this trend suggests.

\section{Supersymmetric contributions}
Now we want to investigate whether low-energy supersymmetry can
reconcile a Higgs mass above the direct experimental limit with a good
$\chi^2$ fit of the electroweak data, in the case of
$\sin^2\theta_{\rm eff}$ near the value obtained from leptonic
asymmetries. Our approach is to discard the measurement of $A^b_{FB}$,
which cannot be reproduced by conventional new physics effects, fix
the Higgs mass above its present limit, and look for supersymmetric
corrections that can fake a very light SM Higgs boson.
As we have discussed in the previous section and as
summarized in fig.~\ref{fig1}, this can be achieved if
the new physics contributions to the $\epsilon$ parameters amount to
shifting $\epsilon_2$ and $\epsilon_3$ down by slightly
more than $1~\sigma$, while leaving $\epsilon_1$ essentially
unchanged.
\begin{figure}
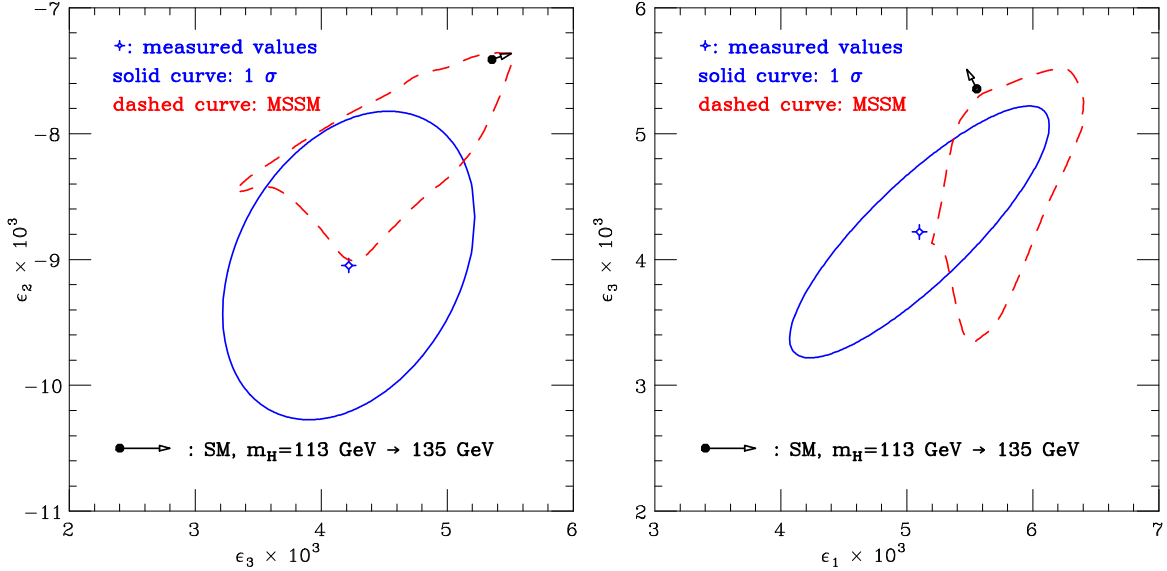

\begin{center}
\epsfig{file=fig2-l.ps,width=0.45\textwidth}
\epsfig{file=fig2-r.ps,width=0.45\textwidth}
\caption{\label{fig2}\it
Measured values (cross) of $\epsilon_3$ and $\epsilon_2$ (left)
and of $\epsilon_1$ and $\epsilon_3$ (right), with their
$1~\sigma$ region (solid ellipses), corresponding to
case {\rm a} of fig.~\ref{fig1}. The area inside the
dashed curves represents the MSSM prediction for $m_{{\tilde e}_L}$ between
96 and 300~{\rm GeV}, $m_{\chi^+}$ between 105 and 300~{\rm GeV},
$-1000$~{\rm GeV}~$<\mu<1000$~{\rm GeV}, $\tan\beta=10$, 
$m_{{\tilde e}_L}=1$~{\rm TeV}. and $m_A=1$~{\rm TeV}.}
\end{center}
\end{figure}

Squark loops cannot induce this kind of shifts in the $\epsilon$
parameters, since their leading effect is a positive contribution to
$\epsilon_1$. Thus, we will assume that all squarks are heavy, with
masses of the order of one TeV. Since the mass of the lightest Higgs
$m_H$ receives a significant contribution from stop loops, we can
treat $m_H$ as an independent parameter and, in our analysis, we fix
$m_H=113$~GeV. Varying the pseudoscalar Higgs mass $m_A$ does not
modify the results of our fit, and therefore we fix $m_A=1$~TeV. The
choice of the right-handed slepton mass has also an insignificant
effect on the fit.
Therefore, we are left with four relevant supersymmetric free
parameters: the weak gaugino mass $M_2$, the higgsino mass $\mu$, the
ratio of the Higgs vacuum expectation values $\tan\beta$ (which are
needed to describe the chargino--neutralino sector), and a
supersymmetry-breaking mass for the left-handed sleptons, ${\tilde
m}_{\ell_L}$ (lepton flavour universality is assumed).
The choice of the $B$-ino mass parameter $M_1$ does
not significantly affect our results and, for simplicity, we have
assumed the gaugino unification relation $M_1=
\frac{5}{3} M_2 \tan^2\theta_W $.

We have computed the supersymmetric one-loop contributions to
$\epsilon_1$, $\epsilon_2$ and $\epsilon_3$ using the results
presented in ref.~\cite{Pierce,MSSMstudies}, and the package
LoopTools~\cite{LoopTools} for the numerical computation of loop
integrals. Figure~\ref{fig2} shows the range of the $\epsilon$
parameters that can be spanned by varying $M_2$, $\mu$, $\tan\beta$,
and ${\tilde m}_{\ell_L}$, consistently with the present experimental
constraints. We have imposed a limit on charged slepton masses of
96~GeV~\cite{Moriond}, on chargino masses of 103~GeV~\cite{Moriond},
and on the cross section for neutralino production
$\sigma(e^+e^-\to\chi_1^0\chi_2^0\to\mu^+\mu^-\slash{E})<0.1$~pb. We
have also required that the supersymmetric contribution to the muon
anomalous magnetic moment, $a_\mu=(g-2)/2$, lie within the range
$0<\delta a_\mu <7.5\times 10^{-9}$.  As apparent from
fig.~\ref{fig2}, light particles in the chargino--neutralino sector
and light left-handed sleptons shift the values of $\epsilon_i$ in the
favoured direction, and by a sufficient amount to obtain a
satisfactory fit.

\begin{figure}
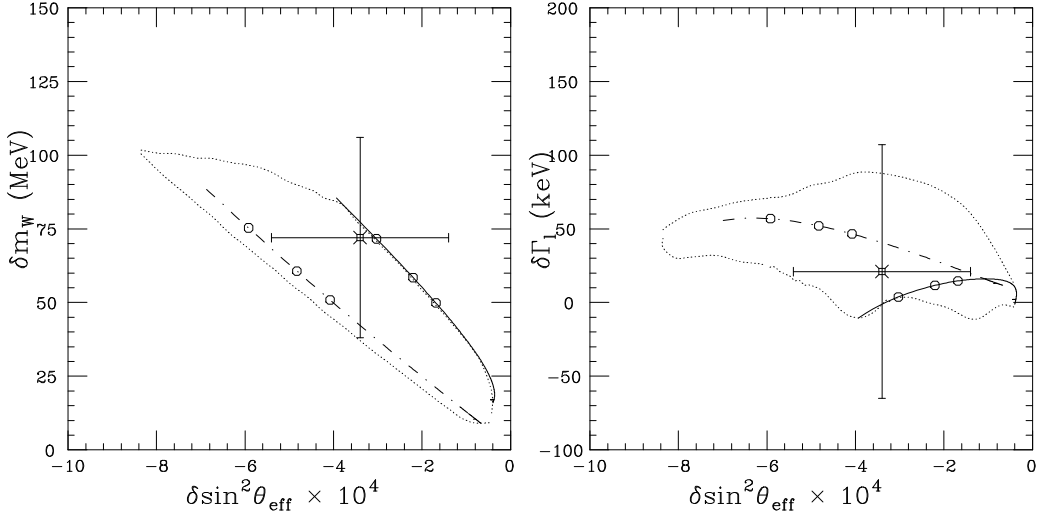

\begin{center}
\epsfig{file=fig3-l.ps,width=0.40\textwidth}
\epsfig{file=fig3-r.ps,width=0.40\textwidth}
\caption{\label{fig3}\it The area inside the dotted curves represents
the shifts in the values of $\sin^2\theta_{\rm eff}$,
$m_W$ and $\Gamma_\ell$ induced by supersymmetric corrections,
for the same parameter region as in fig.~\ref{fig2}. 
The shifts necessary to reproduce the central values of the data
with $m_t=174.3$~{\rm GeV} and $m_H=113$~{\rm GeV} 
are also shown, together with
the corresponding experimental errors. 
The dot-dashed lines are obtained by varying the left slepton
masses, with all other supersymmetric particle decoupled.
The solid curve is obtained analogously, but also keeping a gaugino-like
chargino of $105$~{\rm GeV.}
In each curve, the circles correspond to $m_{\tilde\nu}=60,70,80$~{\rm GeV}
from left to right.
}
\end{center}
\end{figure}

In fig.~\ref{fig3} we show an alternative presentation of our results
directly in terms of the shifts in the observables $m_W$,
$\sin^2\theta_{\rm eff}$ and $\Gamma_\ell$ induced by
supersymmetry.\footnote{A good approximation of the relations
between shifts in the physical observables and in the
$\epsilon$ parameters is given by
$\delta m_W=(0.53\delta\epsilon_1-0.37\delta\epsilon_2-0.32\delta\epsilon_3)
\times 10^5$~MeV; $\delta\Gamma_\ell=(1.01\delta\epsilon_1
-0.22\delta\epsilon_3)\times 10^5$~keV; 
$\delta\sin^2\theta_{\rm eff}=-0.33\delta\epsilon_1+0.43\delta\epsilon_3$.}
For reference, we also display in fig.~\ref{fig3} the difference
between the measured values of the observables (excluding the hadronic
asymmetries) and the corresponding SM predictions for $m_H=113$~GeV,
$m_t=174.3$~GeV. Supersymmetric contributions can bring the
theoretical predictions in perfect agreement with the data.
An interesting observation is that sparticle
effects can increase $m_W$ by $\delta m_W$ up to $\sim 100$~MeV, which
corresponds to approximately three standard deviations, and decrease
$\sin^2\theta_{\rm eff}$ by $\delta \sin^2\theta_{\rm eff}$ up to about
$-8\times 10^{-4}$ ($\sim 4~\sigma$). Note the marked
anticorrelation between $\delta m_W$ and $\delta \sin^2\theta_{\rm eff}$.
$\Gamma_\ell$ is moved upwards, but only by less than 90~keV, or about
$1~\sigma$.

\begin{figure}
\begin{center}
\mbox{\epsfig{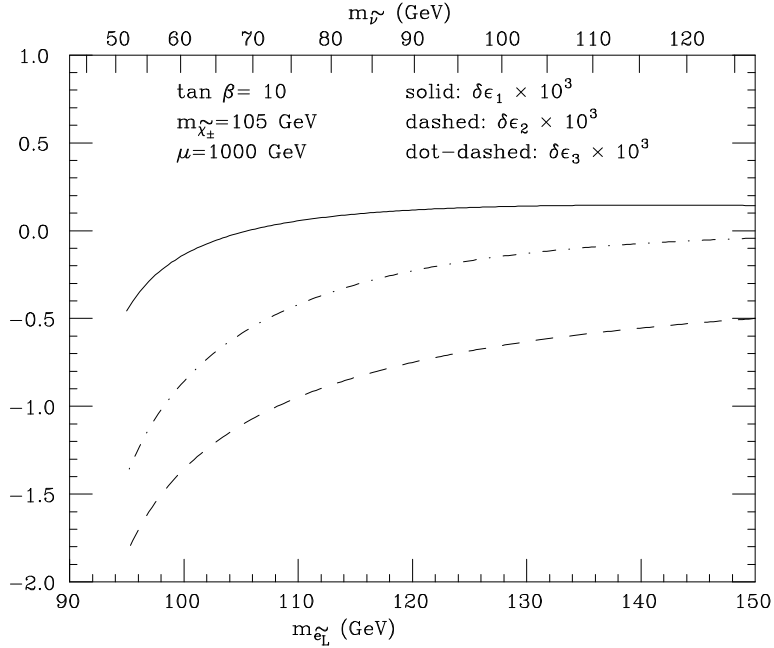}}
\caption{\label{fig4} \it
Supersymmetric contributions to the $\epsilon$ parameters
as functions of the charged slepton (or sneutrino) mass, for a
(purely gaugino) chargino mass of 105~{\rm GeV} and $\tan\beta=10$.}
\end{center}
\end{figure}

\begin{figure}
\begin{center}
\mbox{\epsfig{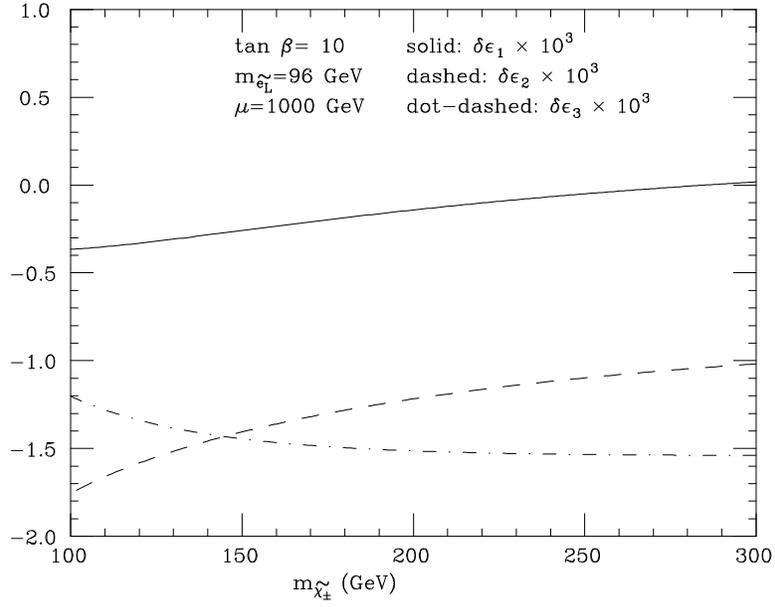}}
\caption{\label{fig5}\it
Supersymmetric contributions to the $\epsilon$ parameters
as functions of the mass of a (purely gaugino) chargino, for a charged
slepton mass of 96~{\rm GeV} and $\tan\beta=10$.}
\end{center}
\end{figure}

\begin{figure}
\begin{center}
\mbox{\epsfig{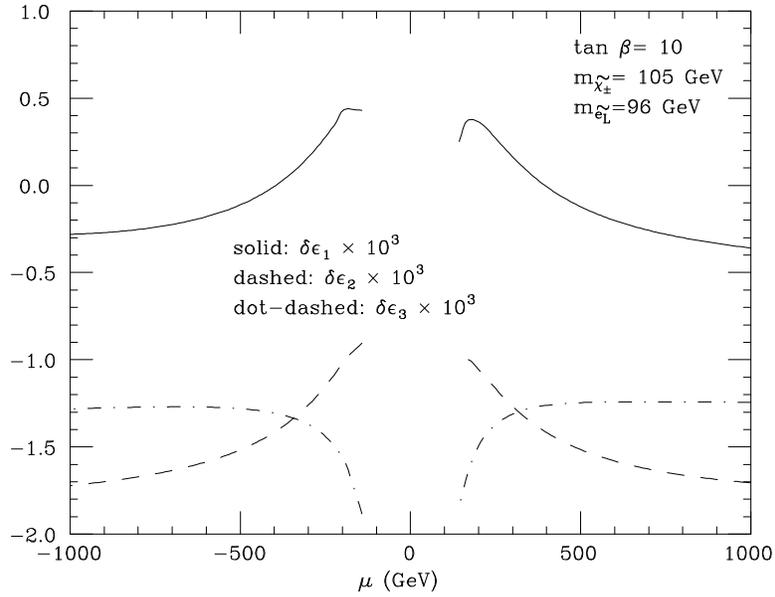}}
\caption{\label{fig6}\it
Supersymmetric contributions to the $\epsilon$ parameters
as functions of the higgsino mass $\mu$, for a charged slepton mass of
96~{\rm GeV}, a chargino mass of 105~{\rm GeV}, and $\tan\beta=10$.}
\end{center}
\end{figure}

Let us now analyse in detail the mass spectrum responsible for this
effects on the $\epsilon$ parameters. The most significant contribution is
coming from light sneutrinos. The effect is maximal when $\tan\beta$ is large
since this allows the smallest possible sneutrino mass compatible with the
charged slepton mass bound,
\beq
m_{{\tilde e}_L}^2=m_{\tilde\nu}^2+m_W^2\left|\cos 2\beta\right|.
\label{slep}
\eeq
Figure~\ref{fig4} shows the supersymmetric contributions to the
$\epsilon$ parameters as functions of the charged slepton (or
sneutrino) mass, for a (purely gaugino) chargino of 105~GeV and for
$\tan\beta=10$. The steep functional dependence of the $\epsilon$'s on
$m_{\tilde\nu}$ illustrates why very light sneutrinos are required to
improve significantly the electroweak fit. The dependence of the
$\epsilon$'s on the lightest chargino mass (again for a purely gaugino
state) is shown in fig.~\ref{fig5}. This dependence is quite milder
than in the sneutrino case. Notice from fig.~\ref{fig5} that, even in
the limit of heavy charginos, in which all the effect is coming from
slepton vacuum polarization contributions, we can obtain a significant
improvement over the SM fit of electroweak data. Light charginos
(mostly because of their contributions to vertex and box diagrams) can
improve the situation, especially by making $|\delta\epsilon_3|
\simlt|\delta\epsilon_2|$, as it seems suggested by the data. Next, we
show in fig.~\ref{fig6} how the supersymmetric contributions to
$\epsilon$'s vary with the lightest chargino composition (or, in other
words, with the parameter $\mu$, for a fixed value of the chargino
mass $m_{\chi^+}=105$~GeV). Part of the region where the lightest chargino is
dominantly a gaugino state ({\it i.e.} large $\mu$) is preferred by
the requirements $\delta\epsilon_1 \simlt 0$ and $|\delta \epsilon_3|
\simlt|\delta\epsilon_2|$, suggested by the data.
For illustration purposes,
the bound $0< \delta a_\mu < 7.5\times 10^{-9}$ has not been imposed in
fig.~\ref{fig6}. It would have the effect of excluding the region of negative
$\mu$, and the region where the lightest chargino is dominantly
a higgsino (small $|\mu|$).


The effect of the light sneutrinos on the electroweak observables is also
explicitly shown in fig.~\ref{fig3}. The dot-dashed lines show the 
contribution of light left sleptons, when all other supersymmetric
particles are decoupled. It is apparent that left sleptons alone are
responsible for the largest part of the effect. When light gauginos
are added to the spectrum (see solid lines of fig.~\ref{fig3})
$\sin^2\theta_{\rm eff}$ increases, $\Gamma_\ell$ decreases,
and $m_W$ remains constant, bringing the theoretical prediction
to an even better agreement with the data. On the other hand, light higgsinos
(which appear only in vacuum polarization diagrams) further decrease
$\sin^2\theta_{\rm eff}$ and increase $m_W$ with respect to the sneutrino
contribution.

To summarize, the request of an improved electroweak data fit is
making precise demands on the supersymmetric mass spectrum. The
left-handed charged sleptons have to be very close to their experimental
bounds, the sneutrino mass is selected to be below about $80$~GeV, the
squarks are in the TeV range, and $\tan\beta \simgt 4$, while there is
no information on right-handed slepton masses. The lightest chargino,
preferably a gaugino state with mass below about 150~GeV, further
improves the fit. This range of supersymmetric parameters is very
adequate in explaining the alleged discrepancy between the
experimental and theoretical values of the muon anomalous magnetic
moment~\cite{gminustwo}. In practice, requiring the supersymmetric contribution
to $g-2$ to be in the range indicated by the data amounts to determining
a precise value of $\tan\beta$ and selecting a
sign (positive in our conventions) of the parameter $\mu$.
We recall that, for moderately large $\tan\beta$,
the negative sign of $\mu$ is disfavoured by
the present measurements of the $B\to X_s \gamma$ branching ratio.

\section{Phenomenological implications}
It is interesting to consider if the requirements obtained in the
previous section on the mass spectrum are consistent with predictions
from the various theoretical schemes proposed for supersymmetry
breaking. Squarks much heavier than sleptons, heavy higgsino states,
and large values of $\tan\beta$ are fairly generic consequences of
supersymmetric models with heavy gluinos and radiative electroweak
symmetry breaking. More unusual is the existence of a sneutrino with
mass less than about $80$~GeV. For instance, in the
supergravity-inspired scheme, in which all sleptons have a common
supersymmetry-breaking mass at the GUT scale and gaugino masses are
unified, we find
\beq
m_{{\tilde e}_L}^2-m_{{\tilde e}_R}^2=0.56M_2^2
+\frac{m_Z^2}{2} (1-4\sin^2\theta_W)\left|\cos 2\beta\right|.
\eeq
This relation, together with eq.~(\ref{slep}), implies that
$m_{\tilde\nu}\simgt m_{{\tilde e}_R}$, once we use the chargino mass limit
$M_2\simgt 100$~GeV.  The experimental limit on $m_{{\tilde e}_R}$
rules out the possibility of a very light sneutrino. Therefore,
$m_{\tilde\nu}<80$~GeV requires different
supersymmetry-breaking masses for left and right sleptons. This could
be achieved in supergravity GUT schemes with non-universal soft
masses,  by giving different scalar mass terms to matter fields
in the $\bf\bar 5$ and in the $\bf 10$
representations of $SU(5)$.
If we call $m_0$ the left slepton soft mass at the GUT scale,
the sneutrino mass is approximately given by
\be
m_{\tilde\nu}^2=m_0^2+0.78 M_2^2-\frac{m_Z^2}{2}\left|\cos 2\beta\right|.
\ee
If we impose $m_0^2>0$, the requirement $m_{\tilde\nu}<80$~GeV
implies $M_2<116$~GeV, and therefore the chargino should lie
just beyond its experimental limit.

Gauge-mediated supersymmetry breaking models~\cite{gaugemed} always predict
$m_{{\tilde r}_R}<m_{{\tilde\ell}_L}$, and exclude the existence of a
very light sneutrino. On the other hand, this is a possibility in
anomaly-mediated models~\cite{anomalymed} with an additional universal
supersymmetry-breaking scalar mass, since the right and left charged
sleptons turn out to be nearly degenerate in mass~\cite{GiudiceWells}:
\beq
m_{{\tilde e}_L}^2-m_{{\tilde e}_R}^2\simeq
0.04\left(m_Z^2\left|\cos 2\beta\right|
+M_2^2 \ln \frac{m_{{\tilde e}_R}}{m_Z}\right).
\eeq
In the case of anomaly mediation, the relation between gaugino masses
is $M_1=11 M_2\tan^2\theta_W$, but this does not give any sizeable
modification of the results shown in figs.~\ref{fig2}--\ref{fig6}. Therefore,
both GUT supergravity schemes with non-universal mass terms and
anomaly mediation can give mass spectra compatible with the requirements
discussed in the previous section.

The selected supersymmetric mass spectrum, with sleptons and possibly
charginos just beyond the present experimental bounds, is certainly
very encouraging for the next generation of experiments. Future hadron
and linear colliders can fully probe this parameter region. However,
the phenomenology may be slightly unconventional. Indeed, the lightest
supersymmetric particle (LSP) is either a neutralino (most probably
a $B$-ino state) or the sneutrino. In gravity mediation with gaugino
unification, the neutralino can be the LSP only if 
$m_{\chi^\pm}<$~110--120~GeV. In anomaly mediation, there is the
possibility that an almost mass-degenerate $SU(2)$ triplet of gaugino 
states is the LSP, and the LEP bound on chargino masses is evaded. Otherwise,
the sneutrino is the LSP.

A light spectrum of electroweak interacting sparticles is promising 
for early discovery.
Hard leptons generated from the decay chains of supersymmetric particles
are the generic signature of our scenario with light sleptons. This is
particularly promising for searches at the Tevatron
that rely on trilepton events. The trilepton topology is
generated by production of a $\chi_1^\pm \chi_2^0$ pair with a subsequent
fully leptonic decay. In our case, we expect that the dominant decay modes
of the ($W$-ino like) next-to-lightest neutralino are $\chi_2^0 \to
{\tilde \ell}_L^\pm \ell^\mp$ and $\chi_2^0 \to
{\tilde\nu}\nu$ and, for the chargino, $\chi_1^\pm \to{\tilde\nu} 
\ell^\pm$ and $\chi_1^\pm \to{\tilde \ell}_L^\pm \nu$.
The decay modes into ${\tilde\ell}_R$ are strongly suppressed in the pure
$W$-ino limit, while an excess of $\tau$ in the final state is present
for a significant gaugino--higgsino mixing.

The slepton decay modes depend on the nature of the
LSP. However in either case (${\tilde \ell}_L^\pm \to \ell^\pm \chi_1^0$
and ${\tilde\nu} \to \nu\chi_1^0$ for $\chi_1^0$ LSP or
${\tilde \ell}_L^\pm \to {\tilde\nu}^\prime \ell^{\pm \prime} \nu$,
${\tilde \ell}_L^\pm \to {\tilde \nu} {\bar f}f^\prime$ 
for ${\tilde \nu}$ LSP),
the final states are rather similar. Notice however that, for a sneutrino
LSP, the charged slepton can decay also into a charged lepton of a different
flavour, ${\tilde \ell}_L^\pm \to \ell^{\pm \prime} \slash{E}$, or into
quarks. The branching ratio into a single
trilepton channel is approximately
\beq
BR( \chi_2^0 \to \mu^+ \mu^- \slash{E})\times 
BR( \chi_1^\pm \to \mu^\pm \slash{E})=\frac{1}{9}
\left[ 1+ \left( \frac{m_{\chi_2^0}^2 -m_{\tilde \nu}^2}
{m_{\chi_2^0}^2 -m_{{\tilde \ell}_L}^2}\right)^2\right]^{-1}.
\label{br}
\eeq
At present the experimental limit on the cross section of a single
trilepton channel is $\sigma(3\mu)<0.05$~pb for 
$m_{\chi_1^\pm}=100$--$120$~GeV~\cite{trilepton}.
Since the cross section for production of gaugino-like $\chi_1^\pm \chi_2^0$
at $\sqrt{s}=2$~TeV is 0.3~pb (0.2~pb) for $m_{\chi_1^\pm,\chi_2^0}=$~100~GeV
(120~GeV), the signal rate (which is obtained by multiplying eq.~(\ref{br})
by the cross section) is not far beyond the present limit, and within
reach of the Tevatron upgrading.\footnote{We thank G. Polesello
for help in the numerical calculation.}

Let us now make some remarks on the relic abundance of the LSP in the
scenario discussed here. Sneutrinos rapidly annihilate with
antisneutrinos in the early universe through $Z^0$ exchange in the
$s$-channel. Even in case of a cosmic lepton asymmetry, their
number density would still be depleted by the process ${\tilde\nu}
{\tilde \nu}\to\nu\nu$ via neutralino $t$-channel exchange. This
annihilation process is efficient, having an $s$-wave contribution,
and it leads to a present sneutrino relic density
\beq
\Omega_{\tilde\nu} h^2 \simeq 10^{-3}\left( \frac{M_2}{100~{\rm GeV}} \right)^2
\left(1+\frac{m_{\tilde \nu}^2}{M_2^2}\right).
\eeq
Values of $\Omega_{\tilde\nu}$ interesting for the dark matter problem
would require $M_2\simgt 1$~TeV. At any rate, since the
sneutrino--nucleon scattering cross section, in the non-relativistic
regime, is 4 times larger than the cross section for a Dirac neutrino
of the same mass, the case of a sneutrino
with halo density in our galaxy has been ruled out by nuclear recoil
detection searches. Nevertheless, it has been
suggested~\cite{snudm} that a cold dark matter sneutrino could be
resurrected in presence of a lepton-number violating interaction that
splits the real and imaginary parts of the sneutrino field, since this
would lead to a vanishing coupling of the LSP to the $Z^0$ boson.

Cosmologically more interesting is the case of a neutralino LSP. For a
$B$-ino LSP and for $m_{{\tilde e}_R}\simlt 2m_{\tilde \nu}$, the
neutralino annihilation rate in the early universe
is dominated by ${\tilde\ell}_R$ exchange, and its 
relic abundance is approximately given by~\cite{Drees}
\beq
\Omega_\chi h^2\simeq \frac{m_{{\tilde e}_R}^4}{{\rm TeV}^2~m_{\chi_1^0}^2}
f\left(\frac{m_{\chi_1^0}^2}{m_{{\tilde e}_R}^2}\right),
\eeq
where $f(x)=(1+x)^4/(1+x^2)$. For instance, for $m_{\chi_1^0}=60$~GeV
and $m_{{\tilde e}_R}=130$~GeV, we obtain $\Omega_\chi=0.3$ (for a
Hubble constant $h=0.7$).  If $m_{{\tilde e}_R}\simgt 2m_{\tilde\nu}$,
then $t$-channel sneutrino and left charged slepton exchange dominate
the annihilation cross section. Since the hypercharge of left sleptons
is half the hypercharge of right sleptons, even in this case we obtain
an appropriate value of the neutralino relic abundance. For instance,
for $m_{\chi_1^0}=60$~GeV and $m_{\tilde\nu}=70$~GeV, we find
$\Omega_\chi=0.5$. However, we recall that coannihilation effects~\cite{coh}
between $\tilde\nu$ and $\chi_1^0$ could significantly reduce
the estimate of the relic abundance given here. Nevertheless,
we can conclude that the supersymmetric mass spectrum
selected by our analysis of electroweak data can predict the correct
$\chi_1^0$ relic density to explain dark matter.

\section{Conclusions}
The long era of precision tests of the SM is now essentially
completed. The result has been a confirmation of the SM to a level
that was hardly believable apriori. In fact, on conceptual grounds, we
expect new physics near the electroweak scale. The fitted Higgs mass
from the radiative corrections is remarkably light. This fact is in
favour of a picture of electroweak symmetry breaking in terms of
fundamental Higgs fields like in supersymmetric extensions of the SM. A light
Higgs in the MSSM should be accompanied by a relatively light spectrum
of sparticles so that it would be natural to expect some of the
lightest supersymmetric particles to be close to their present experimental
limits. Although it is well known that the supersymmetric corrections to the
relevant electroweak observables are rather small for sparticles that
obey present experimental limits, still it is possible that some of
these effects distort the SM quantitative description with shifts of a
magnitude of the order of the present experimental errors. So it is
interesting to look at small discrepancies in the data that could be
attributed to supersymmetric effects. One such effect is the small excess of
the measured value of $m_W$ with respect to the SM prediction for the
observed value of $m_t$ and $m_H$ in agreement with the present direct
lower bound. Alternatively, the same effect is manifested by a
corresponding deficit of the $\epsilon_2$ parameter. Another effect
could be hidden by the fact that unfortunately there is an
experimental discrepancy between the values of $\sin^2\theta_{\rm
  eff}$ measured from leptonic and hadronic asymmetries. If eventually
the true value will be established to be more on the side of the
leptonic asymmetries, then an effect of the same order of that
present in $\epsilon_2$ will also be needed in $\epsilon_3$ to better
reconcile the fitted value of $m_H$ with the direct limits on the
Higgs mass. 

We have shown in this note that negative shifts in
$\epsilon_2$ and $\epsilon_3$ of a comparable size would indeed be
induced by light sleptons and moderately large $\tan\beta$. Charged
slepton near their experimental limit of about $100$~GeV could well
be compatible at large $\tan{\beta}$ with sneutrinos of masses as low
as $55-80$~GeV. If accompanied by light charginos and neutralinos one
can obtain shifts in the radiative corrections of precisely the right
pattern and magnitude to reproduce the description of the data that we
discussed. Light sleptons and large $\tan{\beta}$ are also compatible
with the recent indication of a deviation in the muon $g-2$. We have
discussed the phenomenological implications of this situation. 
Interestingly, the discovery of supersymmetric particles at the Tevatron
in the next few years could be possible in channels with three hard
isolated leptons.

\section*{Acknowledgements}
We thank Anna Lipniacka for useful information, and
Giacomo Polesello for his precious help. 
This work was supported in part
by EU TMR contract FMRX-CT98-0194 (DG 12-MIHT).

\vfill
\end{document}